# Designing Anisotropic Microstructures with Spectral Density Function

Akshay Iyer[1], Rabindra Dulal[2], Yichi Zhang[1], Umar Farooq Ghumman[1], TeYu Chien[2], Ganesh Balasubramanian[3], Wei Chen[1*]

[1]Department of Mechanical Engineering, Northwestern University, Evanston, IL 60208, USA

[2]Department of Physics & Astronomy, University of Wyoming, Laramie, Wyoming 82071, USA

[3]Department of Mechanical Engineering & Mechanics, Lehigh University, Bethlehem, PA 18015, USA

## Abstract

Materials' microstructure strongly influences its performance and is thus a critical aspect in design of functional materials. Previous efforts on microstructure mediated design mostly assume isotropy, which is not ideal when material performance is dependent on an underlying transport phenomenon. In this article, we propose an anisotropic microstructure design strategy that leverages Spectral Density Function (SDF) for rapid reconstruction of high resolution, two phase, isotropic or anisotropic microstructures in 2D and 3D. We demonstrate that SDF microstructure representation provides an intuitive method for quantifying anisotropy through a dimensionless scalar variable termed anisotropy index. The computational efficiency and low dimensional microstructure representation enabled by our method is demonstrated through an active layer design case study for Bulk Heterojunction Organic Photovoltaic Cells (OPVCs). Results indicate that optimized design, exhibiting strong anisotropy, outperforms isotropic active layer designs. Further, we show that Cross-sectional Scanning Tunneling Microscopy and Spectroscopy (XSTM/S) is as an effective tool for characterization of anisotropic microstructures.

## Introduction

Recognizing the pivotal role of microstructure in advanced material systems, microstructure sensitive design [1] has assumed significance in development of advanced material systems. A primary challenge in this domain is the development of quantitative representation of microstructure, ideally involving dimensionality reduction to enable tractable design solutions. Microstructure Characterization and Reconstruction (MCR) aims to capture salient microstructural features and subsequently generate statistically equivalent reconstructions. Bostanabad et al. [2] reviewed existing MCR methods and provide detailed discussions on each. The most well-known MCR method is the Spatial Correlation functions [3, 4], which provide a probabilistic representation of material distribution but, rely on computationally intensive simulated annealing process for reconstruction. Descriptor based method [5, 6] represents microstructures using a small set of uncorrelated descriptors that embody significant microstructural details. Reconstruction involves a hierarchical optimization strategy to match descriptors of reconstructed microstructures to targeted values. However, usage of regular geometrical features and assumption of ellipsoidal clusters deters usage for microstructures with irregular geometries. Machine learning techniques, with their superior capability to learn and reconstruct complex features from isotropic/anisotropic microstructures, have gained popularity as a reconstruction tool. Application of Instance based learning using Support vector machines [7], Supervised Learning [8, 9] and Transfer Learning [10] have shown good reconstruction accuracy, but these methods do not provide a meaningful microstructure characterization, hindering their use in materials design. Deep learning methods such as Convolutional Deep Belief Networks [11] and Generative Adversarial Network [12] can provide a low-dimensional microstructure characterization that could be used as design variables. However, training networks that can be generalized requires several hundred images; which is not readily available for most material systems. Work by Sundararaghavan et al. [13] and Liu et al. [14] displayed capability of Texture Synthesis methods to reconstruct isotropic/anisotropic microstructures from an exemplar, but do not provide a meaningful microstructure representation needed for design.

---

[*] Corresponding Author: weichen@northwestern.edu

Spectral Density Function (SDF) [15-20], a frequency domain microstructure representation, has received a lot of attention for its capability to provide low dimensional, physically meaningful description of quasi-random material systems. For isotropic materials, SDF is one dimensional function of spatial frequency and represents spatial correlations in the frequency domain. Although information contained in SDF is equivalent to two-point autocorrelation function, Yu et al. [15] have shown that SDF provides a more convenient representation for designing microstructures. SDF based microstructure reconstruction can be accomplished in several ways. Chen et al. [21] presented a generalization of Yeong-Torquato reconstruction procedure [3, 4] to define an SDF-based energy function and reconstructed disordered hyperuniform two-phase materials with specified property. However, this reconstruction method incurs large computational time due to its iterative nature. An alternative method based on simulating random fields with a specified SDF, is the Cahn's method [22, 23]. This method was successfully employed for computational design and process development of photonic nanostructures [24]. Although Cahn's Scheme is faster than optimization-based methods for reconstructing structures with specified SDF, the computational cost and time for generating high resolution 3D microstructures remains a challenge. Moreover, all the above-mentioned studies relate to isotropic material systems and there are no instances of design of anisotropic microstructures to the best of our knowledge. This presents a major challenge since anisotropy is highly desired in some material systems, especially where the performance is a manifestation of an underlying transport phenomenon such as Organic Photovoltaic Cells (OPVCs), battery, thermoelectric devices, membranes for water filtration etc. In this article, we examine the case of OPVCs in detail.

OPVCs are promising alternatives to traditional Silicon based solar cells due to several advantages – lightweight, flexibility, low production cost, short payback period [25, 26] etc. but, their large-scale production and commercial usage has been plagued by problems of instability and batch to batch variability [27]. A typical OPVC shown in Fig. 1 consists of an active layer sandwiched between electrodes. There are four key processes taking place in the active layer during energy conversion process: (a) Exciton generation by light absorption; (b) exciton diffusion to donor:acceptor interface; (c) separation of charges from excitons to create electrons and holes and

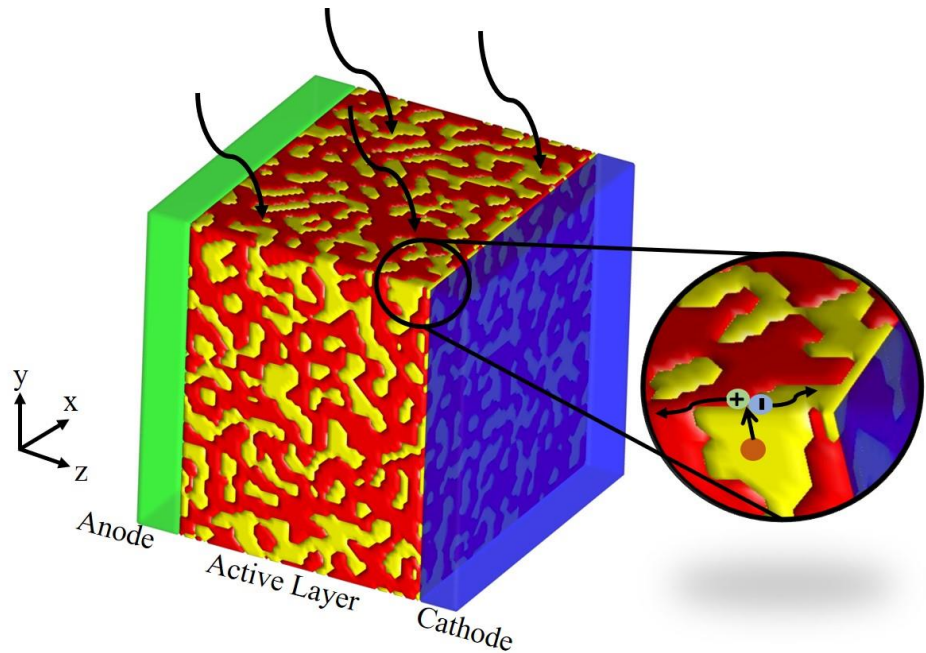

Figure 1: Schematic representation of Organic Photovoltaic cells and the energy conversion process. Magnified scene shows an exciton dissociating at the donor:acceptor interface, followed by charges migrating to respective electrodes.

(d) movement of charges to respective electrodes. Owing to short mean free path of excitons [28], active layers conforming of Bulk Heterojunction (BHJ) architecture is key to ensure high efficiency. The morphology of BHJ, which is controlled by the processing method and related parameters, is crucial in deciding the performance of device [29, 30]. To optimize performance, one would like to maximize the donor:acceptor interfacial area (conversely, minimize the distance an exciton will need to travel) and ensure that the charges can reach their respective electrodes by traversing distance shorter than their mean free path.

In our previous effort [31], we presented a computational framework to establish Processing-Structure-Property (PSP) linkages for OPVC using SDF. We showed that starting from isotropic cross-sectional scanning tunneling microscopy (XSTM) images of spin coated P3HT:PCBM active layer samples, SDF can be used to characterize and reconstruct a 3D Representative Volume Element (RVE) active layer. Leveraging the low dimensional representation enabled by SDF and a novel Structure-Performance relationship to evaluate Incident Photon to Converted Electron (IPCE), we proposed a design formulation with just three variables – two of which arise from SDF parameterization while the third represents the volume ratio of PCBM. Microstructure optimization enhanced performance by 36.75%. Although the optimized design enhanced P3HT:PCBM interfacial area; they do not represent optimal morphology w.r.t charge transport. This is because the tortuosity of PCBM domains forces electrons to traverse longer paths to reach cathode and in some instances, PCBM domains are isolated and do not provide any paths to cathode. Consequently, IPCE is diminished. Intuition dictates that orienting P3HT:PCBM domains in the direction of electrodes i.e. anisotropy will shorten the distance traversed by charges and enhance IPCE.

This article addresses the challenge of anisotropic microstructure design by presenting a novel SDF based reconstruction technique for rapid microstructure generation. Our method is based on Inverse Fourier Transform and can be implemented parsimoniously in any computational package. The method enables reconstruction of anisotropic microstructures without modification. Further, an SDF based anisotropy index is defined to quantify anisotropy and serve as an additional descriptor in the design of strongly anisotropic OPVC active layer that outperforms isotropic designs.

## Results

### Fast Microstructures Reconstruction using Spectral Density Function

SDF based MCR is best suited for quasi-random microstructures that exhibit a seemingly random material distribution but governed by an underlying correlation function. SDF represents spatial correlation in the spatial frequency domain, providing a simplified and physics aware representation of microstructure. Mathematically, SDF ($\rho$) is the squared magnitude of Fourier Transform of a microstructure $\mathcal{M}$:

$$\rho(\mathbf{k}) = |\mathcal{F}[\mathcal{M}]|^2, \tag{1}$$

where $\mathcal{F}[.]$ represents the Fourier Transform operator and $\mathbf{k}$ is a vector representing spatial frequency. A homogeneous microstructure can be described as a realization of an underlying stationary random field and reconstruction involves finding the random field with a prescribed SDF. Towards this end, we propose casting the reconstruction process as a Linear Time Invariant (LTI) system that takes in a random white noise image and transform it into an image (microstructure) with desired SDF using the relationship:

$$\rho_R(\mathbf{k}) = \rho_T(\mathbf{k}) \cdot \rho_W(\mathbf{k}), \tag{2}$$

where . represents point-wise multiplication, subscripts R, T and W denote the SDF of reconstructed, target and white noise images. Note that the white noise image and target image must have the same resolution. SDF is the squared magnitude of Fourier transform and hence, we can recover the reconstructed microstructure from equation (2) by level cutting $\mathcal{M}_R$ to the desired composition of two phases:

$$\mathcal{M}_R = \mathcal{F}^{-1}\left\{\sqrt{\rho_T(\mathbf{k}) \cdot \rho_W(\mathbf{k})}\right\} = \mathcal{F}^{-1}\{|\mathcal{F}\{\mathcal{M}_T\}| \cdot |\mathcal{F}\{\mathcal{M}_W\}|\}. \tag{3}$$

Here, subscripts $R, T\ and\ W$ denote reconstructed, target and white noise microstructures ($\mathcal{M}$) respectively. Since a white noise image contains all frequencies in equal measure, equation (3) can be interpreted as follows: the reconstruction process works like an LTI system with impulse response $\rho_T$ to filter out all frequencies from white noise image except the ones present in $\mathcal{M}_T$. Analogously, reconstruction is a convolution between a white noise image and target image. Fig. 2 shows some examples of microstructures generated using equation (3). Note that $M_R$ will have the same resolution as $\mathcal{M}_T$ (or $\rho_T$). The use of white noise image in equation (3) introduces stochasticity, a key feature across all MCR methods and leads to an ensemble of statistically equivalent reconstructions sharing a common SDF as shown in Fig. 2(A & B).

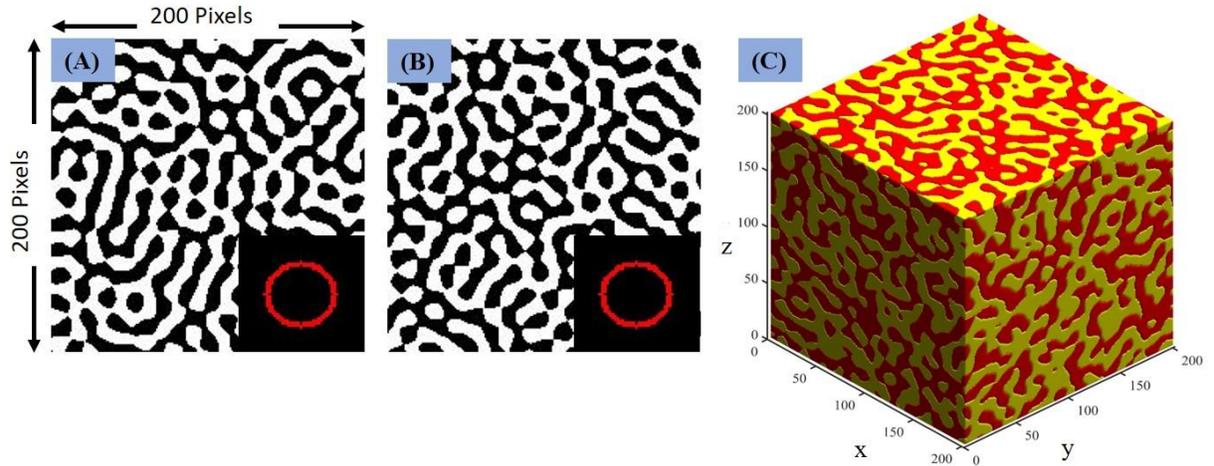

Figure 2: SDF based microstructure reconstruction. (A & B) Two 2D microstructures (200 x 200 pixels) generated from identical ring-type 2D SDFs shown in inset (zero frequency component shifted to center of spectrum). White phases volume fraction is 50% in both cases. In the plot of SDF, the black regions represent frequencies with zero intensity. (C) 3D microstructure (200 x 200 x 200 voxels) generated from an equivalent 3D ring type SDF. Yellow phase volume fraction is 50%.

SDF based reconstruction using equation (3) has two primary advantages as compared to existing methods. First, the reconstruction process only involves Fourier transform and Inverse Fourier Transform, which can be accomplished very efficiently using Fast Fourier Transform available in any computational software package. This is highly significant in terms of generating high resolution 3D microstructures such as those required for investigating OPVCs and performance optimization which will require evaluation of structure-performance model at each iteration. Table 1 compares the Cahn's method [22] against the proposed method w.r.t computational time required for 2D and 3D reconstructions. These tests were performed on a 3.60GHz Intel ® Core™ i7 processor with 24GB RAM. It is quite evident that our method is significantly more efficient than existing methods.

Table 1: Examining computational efficiency of reconstruction methods

| Dimension | Resolution | Computational Time | |
|---|---|---|---|
| | | Our Method | Cahn's Method |
| 2D | 100 x 100 pixels | 0.002 seconds | 0.175 seconds |
| | 200 x 200 pixels | 0.003 seconds | 0.510 seconds |
| | 400 x 400 pixels | 0.008 seconds | 2.254 seconds |
| 3D | 50 x 50 x 50 voxels | 0.039 seconds | 176.511 seconds |
| | 100 x 100 x 100 voxels | 0.087 seconds | 1291.499 seconds |
| | 200 x 200 x 200 voxels | 0.802 seconds | 3.3 hours |
| | 400 x 400 x 400 voxels | 7.337 seconds | Out of Memory |

The second advantage of the proposed method is that no modifications are required for reconstruction of anisotropic microstructures using equation (3) since it is valid for any homogeneous microstructure. This is elaborated in following section.

### Spectral Density Function based Anisotropy Index and Microstructure Reconstruction

Unlike isotropy, which is an absolute state, anisotropy is a relative state and necessitates an appropriate quantitative measure. We define an SDF based anisotropy index with the observation that dominant structural features in spatial domain manifest as non-zero values of frequency components. The pattern formed by these dominant frequencies depends on the microstructural features. Here we discuss two such patterns – ring and disk type SDFs but the concept can be easily generalized to other patterns.

For ring type SDF, we define anisotropy index $\alpha$ as the sine of polar angle $\omega$ subtended by the non-zero frequency component on the axis of anisotropy.

$$\alpha = \sin(\omega), \qquad (4)$$

Fig. 3 shows three sample 2D & 3D microstructures with different degrees of anisotropy, quantified by $\alpha$. As noted earlier, the dimension of SDF is same as that of microstructure. Thus, the 2D microstructures (Fig. 3 A1, B1,C1) have 2D SDFs while 3D microstructures (Fig. 3 A3, B3,C3) have 3D SDFs shown in Fig. 3A2,B2,C2. Microstructures were generated by supplying the corresponding SDFs as $\rho_T(\mathbf{k})$ in equation (3). We observe that SDF's of isotropic microstructures possess symmetry about the center (zero frequency) while anisotropy arises from deteriorating symmetry. The angle $\omega$ captures the extent of symmetry deterioration and thus forms a suitable measure. Equation (4) bounds $\alpha$ in the interval $[0,1]$, where 0 and 1 represent isotropy and absolute anisotropy respectively and facilitates its usage as a bounded design variable in the optimization case study presented in the following section.

Anisotropy Index is, in general, a measure of skewness in SDF pattern and can be extended beyond ring type SDFs shown in Fig.3. Consider the 2D microstructures in Fig. 4 which exhibit disk type SDF patterns. Since there are multiple frequencies present in these microstructures, they appear more disordered than those in Fig. 3 which have only one frequency present. A convenient measure of anisotropy i.e. anisotropy index $\alpha$ for these microstructures would be the eccentricity of SDF patterns. Eccentricity is the ratio of the distance between the foci of the ellipse and its major axis length. Its value ranges from 0 (isotropic) to 1 (strongly anisotropic). These microstructures were reconstructed from Eq.3 by specifying the SDFs shown in inset as $\rho_T(\mathbf{k})$. Thus, our method generates anisotropic microstructures when $\rho_T(\mathbf{k})$ is anisotropic; without any modification of equation (3).

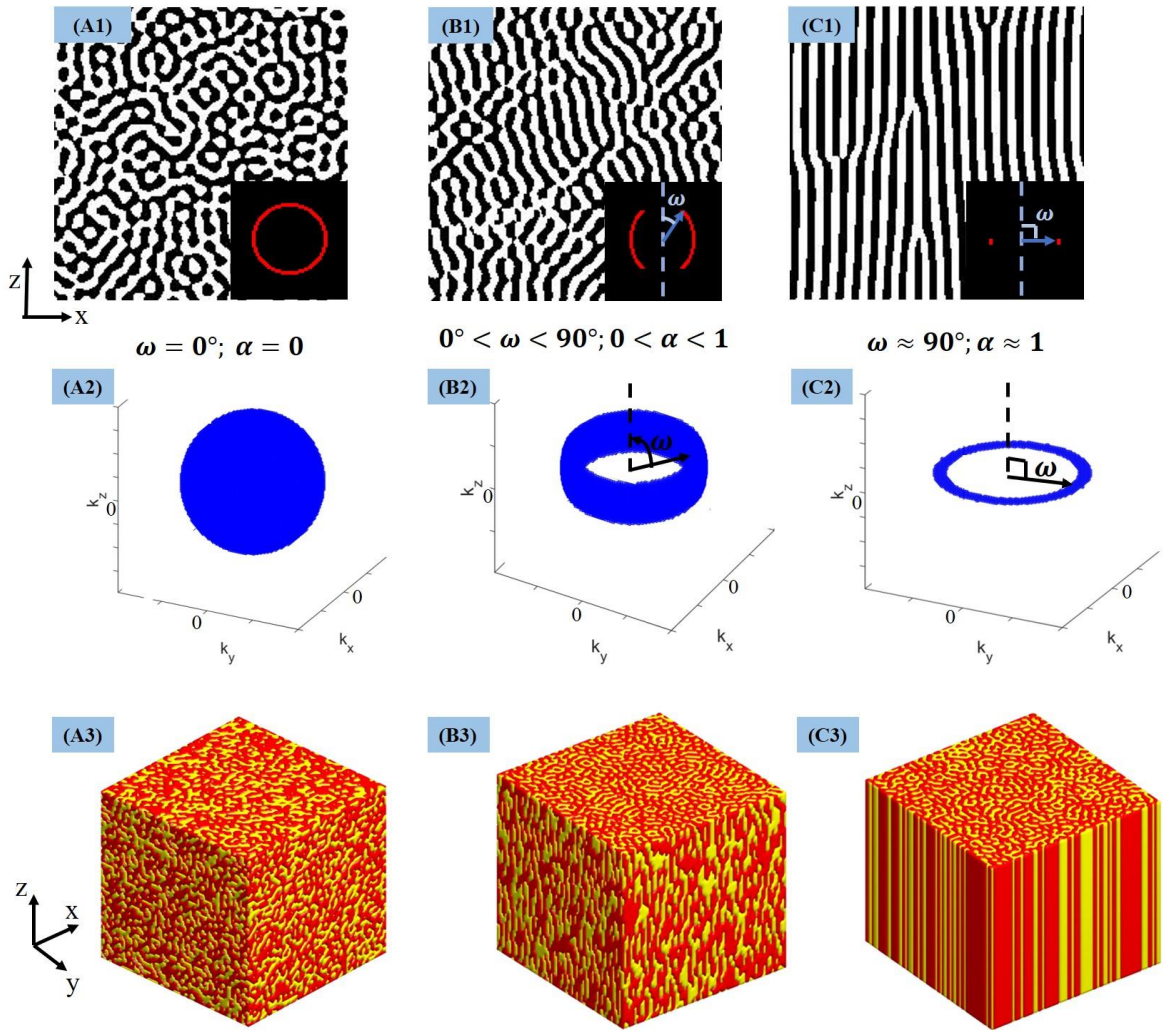

Figure 3: (A1,B1,C1) 2D microstructures with varying degree of anistropy and their SDFs shown in inset. (A3,B3,C3) 3D microstructures and their corresponding SDFs (A2,B2,C2). A1-A3, B1-B3 and C1-C3 represents isotropic, anisotropic and strongly ansiotropic microstructures. Volume fraction of each phase is 50% in all 2D and 3D microstructures shown above.

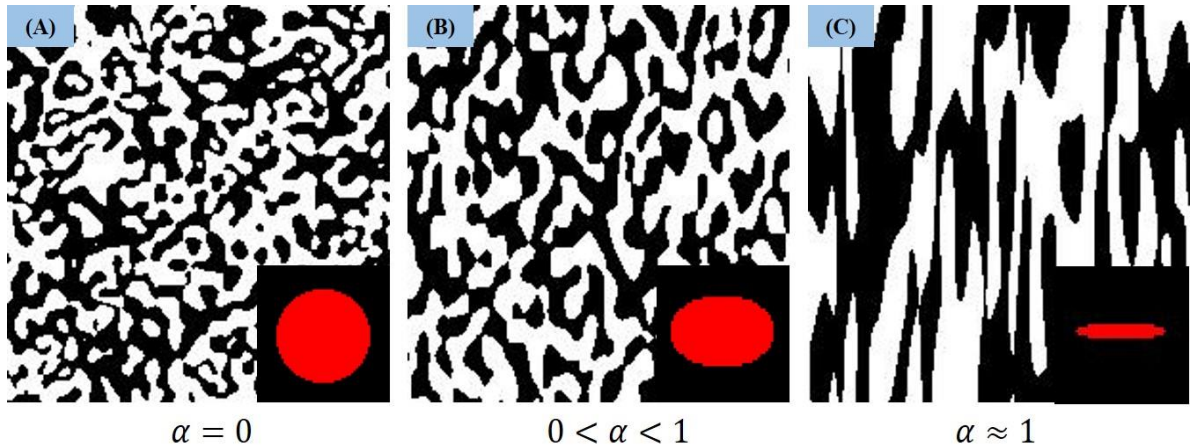

Figure 4: Quantifying anisotropy for micorstructures with elliptical SDF. Each microstructure is 200 x 200 pixels with 50% white phase area fraction. Inset shows corresponding SDFs. Anisotropy index $\alpha$ is defined as eccentricity of SDF pattern.

## Optimizing Active layer Microstructure for OPVCs

As noted earlier, anisotropic active layer can enhance OPVC performance but, realizing such designs has been a challenge hitherto. Here we demonstrate how the SDF based anisotropic microstructure design strategy addresses this challenge.

Given the vast space of possible active layer morphologies, microstructure optimization is necessary to identify active layer designs that maximize Incident photon to Converted electron (IPCE) ratio. The IPCE ratio encapsulates the efficiencies of charge conversion and transport processes which are strongly influenced by active layer microstructure. Ideally, we desire an active layer that maximizes interfacial area and provides short, well-connected pathways for charge transport; but realizing both objectives simultaneously is a challenging task. The low-dimensional microstructure representation enabled by SDF is leveraged to formulate the active layer design as follows:

$$\max_{m \in \mathcal{M}} \text{IPCE}, \tag{5}$$

$$\mathcal{M}: microstructures \ with \ 0\% \leq VF_{PCBM} \leq 100\%, 0.01nm^{-1} \leq k_i \leq 2.23nm^{-1}, 0 \leq \alpha \leq 1$$

where $\mathcal{M}$ is the set of all feasible microstructures, $VF_{PCBM}$ denotes the PCBM concentration by volume ($VF_{PCBM} + VF_{P3HT} = 1$). Here we assume SDF follows a ring type pattern with radius $k_i$, ring thickness $0.01nm^{-1}$ and anisotropy index $\alpha$. $k_i$ controls width of PCBM domains; large values lead to narrower PCBM domains and vice-versa. Bayesian Optimization [32, 33], which adaptively samples designs to efficiently identify global optimum, was applied to solve the formulation presented in equation (5). In each iteration, a high-resolution 3D microstructure (450 x 450 x 450 voxels), embodying a 100nm x 100nm x 100nm Representative Volume Element, is generated using the SDF based reconstruction method discussed previously and its IPCE is evaluated. To show the benefits of anisotropy, we also present results from optimization of isotropic microstructure with design variables $VF_{PCBM}$ and $k_i$ bounded in the range specified by Eq. 5 but $\alpha$ fixed to 0. We performed 100 iterations of Bayesian optimization with expected improvement [34] acquisition criterion.

Fig. 5A indicates that anisotropic designs yield higher IPCE. Table 2 indicates that optimized anisotropic design has an anisotropy index of one; implying that perfect anisotropy is favored. This design is characterized by narrow, wire-like PCBM clusters aligned in the direction of electrodes. These features contribute to a significant reduction in distance traversed by charges to respective electrodes and compensates the reduction in interfacial area. In contrast, the optimized isotropic design maximizes interfacial area, but its effect is subdued by tortuous paths for charge transport, leading to lower IPCE.

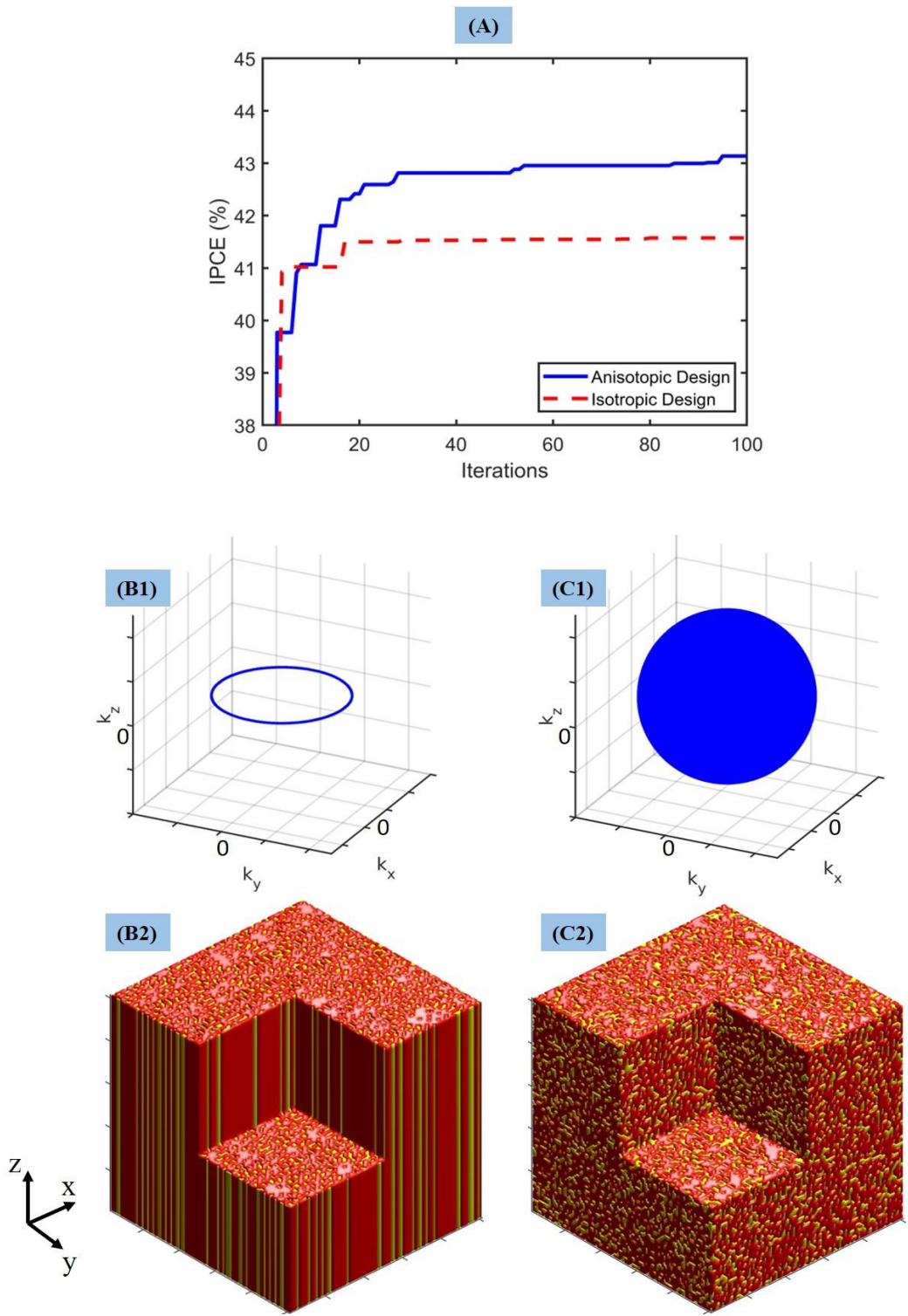

Figure 5: (A) Bayesian Optimization history for IPCE maximization, depicting superior perfomance of anisotropic design as compared to isotropic design. (B1) SDF of Optimized anistropic microstructure (B2) displaying perfect anisotropy along Z – direction. (C1) SDF of Optimized istropic microstructure (C2). Red and yellow phase represents P3HT & PCBM respectively.

Fig. 5(B2 & C2) shows 200 x 200 x 200 voxel segments from the optimized anisotropic and isotropic microstructures and their corresponding SDFs. Anisotropic design, which is isotropic in XY plane but exhibits strong anisotropy along Z-axis, leads to an IPCE of 43.14% as compared to 41.57% of isotropic design. As suggested by Table 2, the difference in IPCE can be attributed to drastic reduction in distance traversed by electron to cathodes ($S_C$), a consequence of strong anisotropy of PCBM domains. The reduction in $S_C$ dominates the minor increase in distance to nearest interface ($d$) observed in anisotropic design w.r.t isotropic design.

Table 2: Optimum design variables and resulting microstructural features

|  |  | Optimized Anisotropic Design | Optimized Isotropic Design |
|---|---|---|---|
| **Optimum Design Variables** | $VF_{PCBM}$ | 0.228 | 0.290 |
|  | $k_i$ | $1.43\ nm^{-1}$ | $1.77\ nm^{-1}$ |
|  | $\alpha$ | 1 | 0 |
| **Microstructural Features** | Average $d$ | $0.26\ nm$ | $0.23\ nm$ |
|  | Average $S_A$ | $50.00\ nm$ | $50.01\ nm$ |
|  | Average $S_C$ | $50.00\ nm$ | $60.93\ nm$ |

## Electric Field Induced Anisotropy in PCBM:P3HT Active Layer

In a P3HT:PCBM mixture, P3HT is the polar molecule. Thus, under the application of electric field the orientations of P3HT-rich domains are affected, leading to elongation of domains. So far, the electric field treatment during the OPVC synthesis processes has only been discussed regarding the better polar molecule crystallinity and change of surface roughness for better contact with electrodes [35-39]. There's a lack of studies depicting the change in active layer nano-morphology of the molecular domains under the influence of an electric field. To address this issue, we examined the effect of electric field treatment induced anisotropy in OPVC active layer using cross-sectional scanning tunneling microscopy and spectroscopy (XSTM/s). This is a novel approach for investigating the effect of electric filed on OPVC at atomic level [40, 41]. A P3HT:PCBM mixture was spin coated onto a Silicon substrate and subjected to electric field during annealing process as illustrated in Fig. 6A. The d$I$/d$V$ mappings of the active layers obtained from this procedure exhibit a clear distinction between the domain textures with (Fig. 6C) and without (Fig. 6B) application of electric field. P3HT domains are elongated in the direction of applied electric field (along Z-axis) while in its absence, these domains have random orientation and render an isotropic texture to morphology (Fig. 6B). These differences are also reflected in their corresponding SDFs; SDF of an isotropic morphology is symmetric around the zero frequency while anisotropic morphologies lack this symmetry. Recent findings by Dulal et al. [41] show that in addition to influencing film morphology, application of electric field alters electronic properties of P3HT & PCBM domains due to molecular intermixing induced doping.

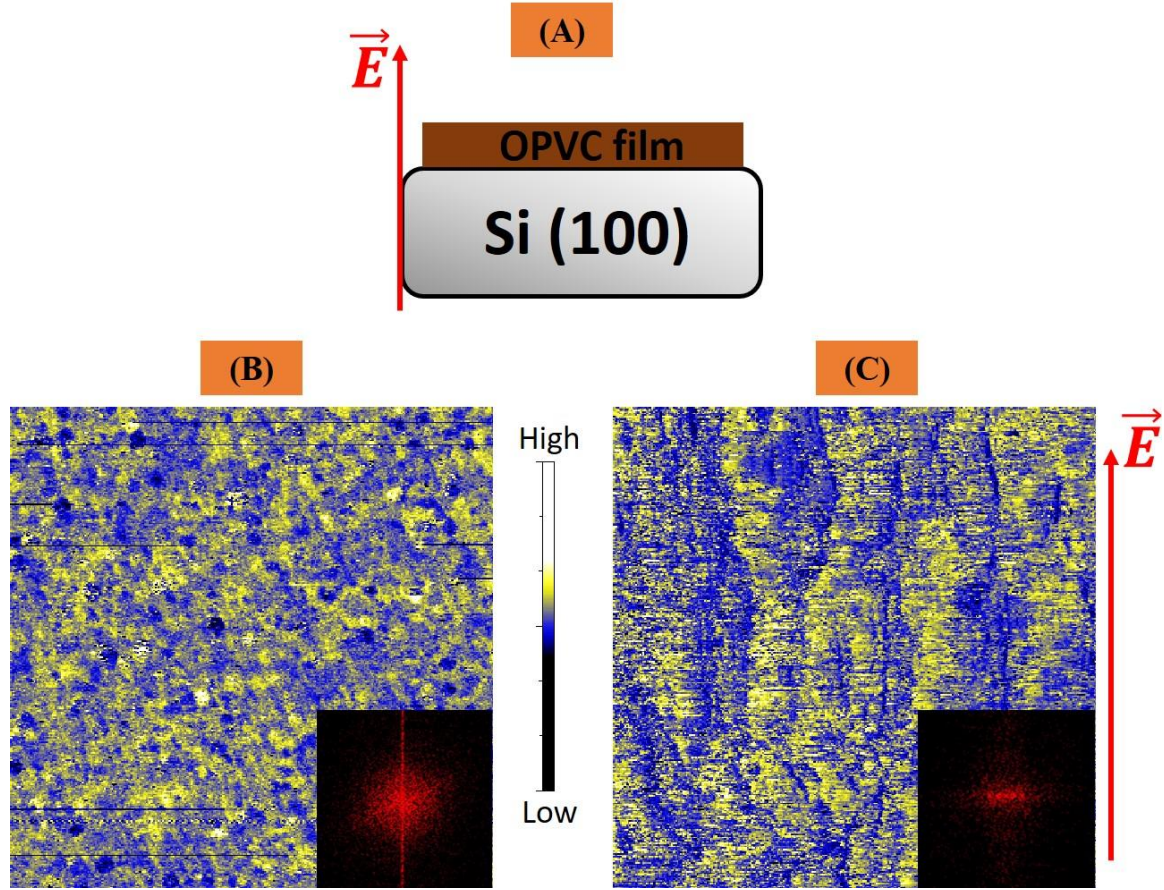

Figure 6: (A) Experimental setup showing the sample geometry and the direction of the application of the Electric field for the OPVC active layer from side view. (B) A 160 nm × 160 nm XSTM/S d$I$/d$V$ mapping of P3HT:PCBM film not subjected to electric field. (C) A 160 nm × 160 nm d$I$/d$V$ mapping of P3HT:PCBM film subjected to electric field during annealing. Direction of electric field shown by arrow. Yellow and blue domains correspond to P3HT, PCBM respectively. SDF for each dI/dV mapping shown in corresponding inset.

## Discussion

The new SDF based microstructure reconstruction methodology proposed is capable of fast reconstructing high resolution, isotropic/anisotropic microstructures which enables computational microstructure design in a wide range of materials systems. The representation is physics-aware and highly flexible for both isotropic and anisotropic microstructures with a small set of parameters that can serve as design variables in microstructure design. To quantify the level of anisotropy, we introduced anisotropy index, a scalar value bound in the interval [0,1] with 0 and 1 representing purely isotropic & anisotropic microstructures respectively. Meeting the need to reduce charge transport distances in the active layer, we formulated a design case study that pivots on SDF to find the optimal active layer microstructure with an aim to maximize IPCE. Only three design variables – PCBM volume fraction, anisotropy index and one SDF parameter were sufficient to represent microstructure, thus making optimization tractable. Optimization results reinforced our intuition that microstructures exhibiting strong anisotropy (anisotropic index ~ 1) in the direction of electrodes provide shorter paths for charges to travel towards respective electrodes; thus, delivering an enhancement in IPCE as compared to isotropic microstructures. STM imaging of P3HT:PCBM films visually confirms the presence of elongated features in the direction of electric field and subsequently reflected by asymmetry in its SDF pattern.

The design strategy for anisotropic microstructures proposed in this article can be extended to design any material systems where anisotropy could potentially enhance performance. Challenges in fabrication of strongly anisotropic morphologies with desired domain size and distribution represent opportunities for future work. In this regard, recent work by Jiali et al. [42] shows that controlled solvent vapor treatment on a highly oriented polyethylene substrate improves crystallization ability and induces strong anisotropy in P3HT films. A holistic design approach requires Processing-Structure-Performance relationships for OPVCs, which is part of our ongoing initiative [43, 44] that employs Coarse Grained Molecular Dynamics to understand Processing-Structure relationship and integrate models across the processing-structure-performance chain using SDF based microstructure representation.

## Methods
### Structure-Performance Simulation

In OPVCs, the energy conversion is accomplished in four main processes: (1) exciton creation by light absorption; (2) diffusion of these excitons; (3) separation of charge at interface; (4) diffusion of charge and its collection at the electrodes. The efficiency of this process is denoted Incident Photon to Converted Electron (IPCE) ratio. Given the active layer microstructure, IPCE can be evaluated using the following expression [31] :

$$IPCE = \frac{1}{A} \ldots$$

(6)

$$\sum \left( e^{-(t-z)\alpha(\lambda)} \Delta x \Delta y \left(1 - e^{-\alpha(\lambda)\Delta z}\right) \right) \left( e^{-\frac{d}{\xi_{ex}}} \right) (P_{sep}) \left( e^{-\frac{S_A}{\xi_h}} e^{-\frac{S_C}{\xi_e}} P_{col} \right),$$

The four parenthesis in above equation represent the four processes mentioned above, where the summation collects contribution of each voxel of the 3D structure; $t$ is the thickness of the active layer (100nm in our case); $z$ is the height of the specific voxel; $\alpha(\lambda)$ is the absorption coefficient of active layer as function of the light wavelength, $\lambda$; $P$ refers to probability for charge separation (sep), and for charge collection (col); $d$ is the distance to the nearest interface from the location of the exciton creation; $\xi$ the diffusion lengths of exciton (ex); of hole (h); and of electron (e); $S_A$ is the shortest distance to anode through the donor phase; and $S_C$ to cathode through the acceptor phase.

For implementation purposes, we make the following assumptions: (i) incident light is along the Z-axis (see Fig. 5), (ii) there are no voids, impurities in active layer microstructure and (iii) $P_{sep}$ and $P_{col}$ are assumed to be 1. Light wavelength $\lambda$ is set to 510 nm, at which P3HT & PCBM absorption coefficients are $1.75 \times 10^7\ m^{-1}$ [45] & $0.21 \times 10^7\ m^{-1}$ [46].

There are four main structural variables viz., $z, d, S_A$ and $S_C$ that must be evaluated at each voxel. The four steps in energy conversion process are simulated as such: (1) First, excitons are created at each voxel with the light intensity determined by the depth of this cell: $t$-$z$. Once the light reaches the bottom, it is reflected back up with its intensity decaying at the same rate; (2) the exciton then diffuses to the nearest interface with a distance $d$; (3) the exciton dissociates into electron and hole at the interface; (4) the hole diffuses towards the anode through the shortest path, $S_A$, in P3HT, while the electron diffuses towards cathode through the shortest path, $S_C$, in PCBM.

P3HT:PCBM Sample Preparation and Imaging
P3HT (regio-regular (RR 93-95) SOL4106, used as received from Solaris Chem Inc.) and PCBM (purity >99.5%, SOL5061, used as received from Solaris Chem Inc.) were first made with into separate solution in chlorobenzene (purity ≥99.5%, sigma-aldrich). Solutions with desired P3HT:PCBM weight ratio 1:1 were made by mixing the precursor solutions. The solutions were then spin coated onto the Si (100) substrate with 1050 rpm speed for 1 minute. The films containing P3HT:PCBM/Si (100) were annealed at 100°C for 20 minutes in inert environment for the sample without electric field. For the sample with electric field treatment, P3HT:PCBM/Si (100) were annealed at 100°C for 20 minutes under the application of electric field of approximately (3.89 ± 0.04 kV/cm) in inert environment. The sample is then mounted on a special sample plate called cross-sectional sample plate for proper clamping and fracturing in UHV environment [40]. The sample is then transferred into STM chamber quickly and fractured in UHV from the side so that the film region is not affected. Finally, the cross-sectional fresh fractured Si and P3HT: PCBM film region is scanned inside the scanning chamber of STM.

Scanning tunneling microscopy and spectroscopy (STM/S) is a special technique based upon quantum tunneling that can obtain topographic images and d$I$/d$V$ mapping [41]. The d$I$/d$V$ mapping also known as the Local Density of States (LDOS) mapping gives the electronic density of states. Various technique like AFM, SEM and TEM are used for imaging the OPVC but those techniques doesn't clearly show the domains of OPVC [47]. The d$I$/d$V$ mapping in STM is the molecular sensitive technique which can image the interfaces of heterojunctions of OPVC showing two domains of OPVC with two contrast region [31]. Cross-sectional scanning tunneling microscopy and spectroscopy (XSTM/S) has been a novel technique to image OPVCs [31] and perovskite solar cells [48]. Cleaving and fracturing of the sample were applied to generate the fresh surface on Si for measurements as organic solar cell active layer degrade upon sputtering and annealing of the active layer.

## Acknowledgement
Authors acknowledge grant support from National Science Foundation (NSF) CMMI-1662435, 1662509 and 1753770 under the Design of Engineering Material Systems (DEMS) program, as well as the support from NSF EEC 1530734. We thank Dr. Ramin Bostanabad for providing the 3D microstructures visualization code.

## Author Contributions
A.I. and W.C. conceived the project. Y.Z. conceived the fast SDF based reconstruction technique and A.I. developed the anisotropy index, prepared the microstructure design formulation and performed optimization for OPVC design case study. U.F.G. provided the Structure–Performance simulation code for the design case study. R.D. fabricated isotropic & anisotropic films under guidance of T.C. G.B. also provided supervisory role in this collaborative project sponsored by NSF.  All authors analyzed the results and contributed in preparation of this article.

## Code Availability

The code for reconstruction technique reported in this paper is available from the corresponding authors upon request.